\documentclass[a4paper]{article}

\usepackage{INTERSPEECH2022}
\usepackage[table,xcdraw]{xcolor}
\usepackage{float}
\usepackage{enumitem}
\usepackage{graphicx}
\usepackage{arydshln}
\usepackage{multirow}
\usepackage{cite}
\usepackage{url}
\title{Interpretable Acoustic Representation Learning on Breathing \\ and Speech Signals for COVID-19 Detection}
\name{Debottam Dutta$^1$, Debarpan Bhattacharya$^1$, Sriram Ganapathy$^1$, \\
Amir H. Poorjam$^2$, Deepak Mittal$^2$,  Maneesh Singh$^2$.\thanks{This work was supported by the grants from Verisk Analytics Inc.}}
\address{
  $^1$LEAP Lab, Indian Institute of Science, Bangalore, India.\\
  $^2$Verisk Analytics, Inc., Jersey City, NJ, USA}
\email{sriramg@iisc.ac.in}

\begin{document}

\maketitle
\begin{abstract}
In this paper, we describe an approach for representation learning of audio signals for the task of COVID-19 detection. The raw audio samples are processed with a bank of 1-D convolutional filters that are parameterized as cosine modulated Gaussian functions. The choice of these kernels allows the interpretation of the filterbanks as smooth band-pass filters. The filtered outputs are pooled, log-compressed and used in a self-attention based relevance weighting mechanism. The relevance weighting emphasizes the key regions of the time-frequency decomposition that are important for the downstream task. The subsequent layers of the model consist of a recurrent architecture and the models are trained for a COVID-19 detection task. In our experiments on the Coswara data set, we show that the proposed model achieves significant performance improvements over the baseline system as well as other representation learning approaches. Further, the approach proposed is shown to be uniformly applicable for speech and breathing signals and for transfer learning from a larger data set. 
\end{abstract}

\section{Introduction}
The biomedical applications, like automatic diagnosis of diseases using speech and audio, have a strong requirement to explain away the basis the model is using to arrive at a specific decision about a sample~\cite{fagherazzi2021voice, avila2021investigating,holzinger2018machine}. 
The early approaches to facilitate this need 
involved the design of knowledge-driven features~\cite{low2020uncovering,gupta2016pathological}.
A wide range of acoustic features that can reflect different voice disorders have been designed, such as jitter, which indicates the instability in the fundamental frequency~\cite{ozdas2004investigation}, shimmer, which is a measure of deviations in amplitude~\cite{alghowinem2013detecting}, harmonic-to-noise ratio, which is an indication of hoarseness in voice~\cite{shama2006study}, and maximum phonation time that indicates the lung capacity~\cite{maslan2011maximum}. 
However, these approaches require extensive experiments to understand their impact on the downstream application. 

In the recent years, with advancements in the deep learning approaches, hand-crafted feature extraction algorithms have been increasingly replaced by data-driven methods. 
This field of study, called representation learning \cite{bengio2013representation}, is the one in which a deep neural network learns representations from raw data without any prior assumptions.
Representation learning is a well-explored field in computer vision~\cite{radford2015unsupervised} and natural language processing~\cite{mikolov2013efficient}, where the representation of the data is learned by directly feeding pixels or one-hot encoded words to the neural network, respectively. 
Recently, there has also been a great interest in learning general purpose representations for speech and audio~\cite{turian2022hear, baevski2020wav2vec}. 
However, the high temporal variability and the curse of dimensionality of the acoustic signals, makes the representation learning challenging \cite{freitag2017audeep}.
In this paper, we explore representation learning for speech directly from
the raw-waveform using a modeling approach that consists of a parametric filter learning module and a self-attention style relevance weighting module. The representation learning approach pursued in this work aims at developing robust front-ends for the task of detecting COVID-19 from audio sounds of speech and breathing nature.  

In audio processing, the most common approach for representation learning has been the direction of learning the time-frequency decomposition from the raw audio.  These approaches can be categorized as supervised and unsupervised methods.
% time-frequency decomposition of the signal.
Examples for supervised learning include phoneme classification \cite{sainath2013learning}, acoustic modeling for speech recognition \cite{hoshen2015speech}, acoustic event classification using learnable audio front-end (LEAF) \cite{zeghidour2021leaf} and music genre classification using deep scattering transform \cite{anden2014deep}. Using parameterized sinc functions, Ravanelli and Bengio proposed a SincNet architecture \cite{ravanelli2018interpretable} in which the first layer of the convolutional neural network (CNN) learns meaningful filters directly from the raw audio signals. 

\begin{figure}[t!]
    \centering
    \includegraphics[width=.47\textwidth]{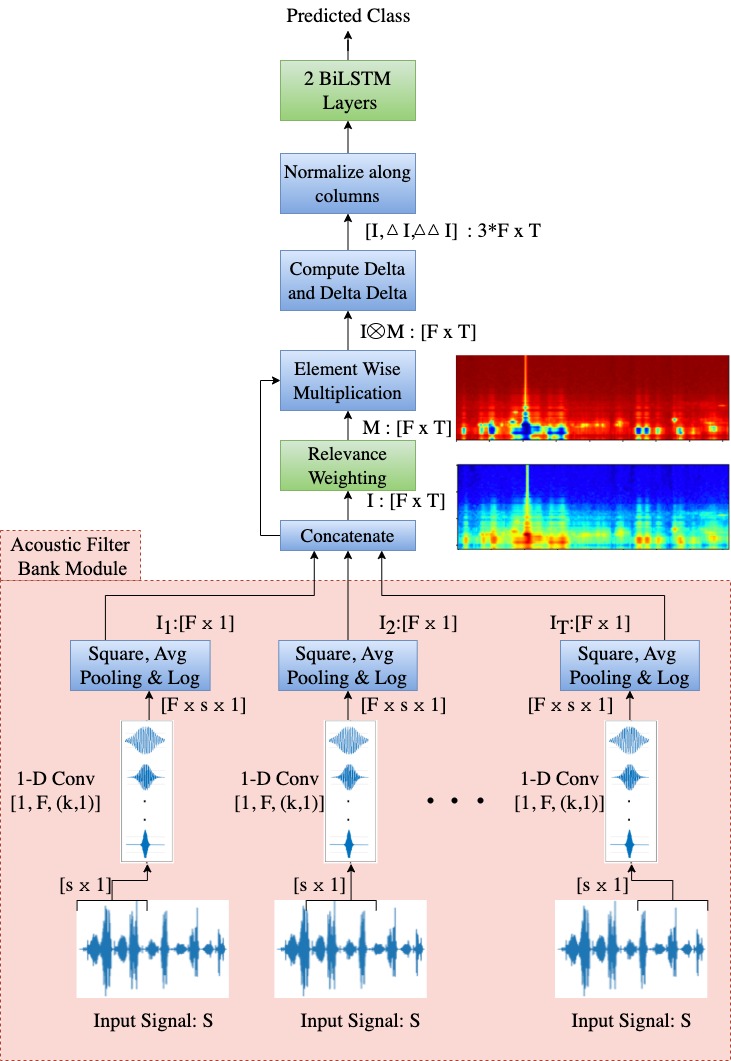}
    \caption{Overall pipeline for filter learning. The input signal is passed through a 1-D convolutional layer where we use Gaussian set of filters. The relevance weighting module consists of a fully connected layer followed by a sigmoid function. Two maps are also shown as output of convolution operation and the corresponding learned mask. Blocks in green depict a learnable function whereas the blue blocks depict fixed functions.} 
    \label{fig:pipeline}
\end{figure}
% \begin{figure}[t!]
%     \centering
%     \includegraphics[width=.45\textwidth]{LaTeX/figs/Interspeech.jpg}
%     \caption{Overall Pipeline for Filter Learning. Input signal is passed through a 1-D convlutional layer where we use gaussian set of filters. Relevance weighting module consists of a fully connected layer follwed by a sigmoid function. Two maps are also shown as output of Convolution operation and then masked Input. Blocks in Green depict a leanable function whereas blue block depict non-learnable functions.} 
%     \label{fig:pipeline}
% \end{figure}
For unsupervised learning of audio representations, the use of restricted Boltzmann machines \cite{agrawal2017unsupervised,sailor2016filterbank} and variational learning \cite{agrawal2019unsupervised} have been explored. The wav2vec method proposed in \cite{schneider2019wav2vec} explored self-supervised pre-training using an autoregressive loss function. Further, the  generative adversarial networks \cite{agrawal2018comparison} have been explored for acoustic/modulation filter learning.

The approach of self-supervised learning  from a number of tasks has been explored by Pascual et al.~\cite{pascual2019learning}.  
In these applications, while representation learning has achieved performance improvements for the downstream tasks considered,  they have been limited in terms of the interpretability.  
The key exception is the SincNet filterbank \cite{ravanelli2018interpretable}  where 
the authors used parametric sinc filters to improve interpretability. 

In this work, we extend the previous work by Agrawal et al.~\cite{agrawal2020interpretable} on filterbank learning using cosine modulated Gaussian filterbanks. We explore the learning of representations on breathing and speech signals for the task of COVID-19 detection.
The raw audio samples are processed through a bank of 1-D convolutional filters. 
A self-attention based sub-band weighting approach, called relevance weighting, provides rich representations of the audio signal for the downstream recurrent neural network layers. 
The key contributions from the proposed work over the previous work in \cite{agrawal2020interpretable} can be summarized as follows, 
 
\begin{itemize}[leftmargin=*]
    \item A unified filter learning paradigm that provides meaningful representations from both speech and breathing signals.
    \item A relevance weighting mechanism that can dynamically emphasize the regions of the time-frequency decomposition that are useful for the task. 
    \item Transfer learning of the filter  representations from a large scale data set of speech/breathing sounds to the final target data set. 
    \item Exploring supervised and self-supervised representation learning  objectives for acoustic filter learning. 
    \item Experimental evidences and analysis to highlight the advantages of the proposed representation learning framework.

\end{itemize}

% It should be noted that since the focus of this paper is on the front-end representation, we fixed the classifier in all experiments to investigate the impact of the proposed representation on the performance.

% \input{LaTeX/figs/fig_cpc}
\section{Acoustic Filterbank with Relevance Weighting}
The block schematic of the proposed front-end is shown in Figure~\ref{fig:pipeline}. In the following subsections, we explain the main components of the proposed method.

\subsection{Acoustic Filterbank Layer}
The acoustic filterbank layer receives raw audio samples which are windowed into a sequence of $s$ samples. 
The windowed samples are processed using a bank of 1-D convolutional filters which are parameterized as a cosine modulated Gaussian function \cite{agrawal2020interpretable}, where the kernels $g_i(n)$, corresponding to the $i$th filter, are denoted as 
\begin{equation}
    g_i(n) = \cos{2\pi \mu_i n} \times \exp{(-n^2\mu_i^2/2)}, 
\end{equation} 
where $(i = 1,2,\ldots,F)$ and $\mu_i$ is the centre frequency of the  \textit{i}-th kernel. Here, $F$ denotes the number of acoustic filters. The  Gaussian function, given in the above equation, represents a low-pass filter while the cosine modulation makes the kernel band-pass. The center frequencies $\mu_i$ are the only learnable parameters of the filter. The band-width of the filters is directly proportional to the center frequency (inverse of the variance in the time-domain kernel function in the above equation). This is to enable the learning of constant-Q filters.    

The given audio signal is first segmented into short-time frames of $s$ samples. 
Each frame is then convolved with the $F$ filters. 
Finally, the output of the filters are squared, average pooled and log-transformed. 
This processing module generates an $F$ dimensional output for each frame. Further, splicing the frame-level $F$ dimensional representations from all $T$ frames of the given audio file, gives the learned time-frequency representation, $\boldsymbol{I}$, of dimension $F\times T$. 
\begin{table*}[tb]
\centering
\setlength{\tabcolsep}{3pt}
\caption{Comparison of the performance of the COVID-19 detection models using different signal representations on breathing and speech modalities in terms of the AUC(\%).}
\vspace{-2mm}
\begin{tabular}{l|cccccc||cccccc}
\hline
% \small
\multirow{2}{*}{Representation Methods} & \multicolumn{6}{c||}{Breathing} & \multicolumn{6}{c}{Speech} \\
\cline{2-13}
 & \multicolumn{1}{l}{Fold1} & \multicolumn{1}{l}{Fold2} & \multicolumn{1}{l}{Fold3} & \multicolumn{1}{l}{Fold4} & \multicolumn{1}{l}{Fold5} & \multicolumn{1}{l||}{Avg.} 
& \multicolumn{1}{l}{Fold1} & \multicolumn{1}{l}{Fold2} & \multicolumn{1}{l}{Fold3} & \multicolumn{1}{l}{Fold4} & \multicolumn{1}{l}{Fold5} & \multicolumn{1}{l}{Avg.}\\ \hline
Mel-Spectrogram~\cite{sharma2021second} & 73.8 & 76.4 & 75.6 & 77.1 & 84.2 & 77.4 & 
74.8 & 85.1 & 79.1 & 78.6 & 80.4 & 79.6 \\
SincNet~\cite{ravanelli2018interpretable} & 72.4 & 69.6 & \textbf{81.9} & 76.4 & 77.1 & 75.4 
& 75.0 & 84.6 & 77.4 & 76.7 & \textbf{85.2} & 79.7 \\
LEAF~\cite{zeghidour2021leaf} & 71.5 & 62.9 & 66.1  & 66.6  & 73.6  & 68.2  & 60.5 & 59.1 & 63.8 & 57.4 & 62.0 & 60.6 \\
%&  &  &   &   &   &   & \\
%CosGauss-relev-2head~\cite{agrawal2020interpretable} & 76.4 & 76.3 & 79.4 & 77.9 & 81.5 & 78.3
 %& 77.1 & 83.3 & 81.5 & 78.4 & 85.7 & 81.2 \\
\hdashline[1pt/1pt]
CosGauss-relev & 75.4 & 76.0 & 79.7 & 78.1 & 82.8 & 78.4 
& 76.5 & \textbf{83.6} & 83.3 & \textbf{79.6} & 84.5 & \textbf{81.5}\\
\hdashline[1pt/1pt]
%CosGauss-relev-modFB &  72.2 & 70.4 &  79.4 &  76.7 & 81.9 & 76.1  
%&    &   &   &   &   &  \\
%CosGauss-relev-modFB-relev & 73.7 & 76.0 & 81.1 & 78.3 & 82.9 & 78.4 
%& 75.6 & 79.4 & 80.7 & 81.2 & 82.6 & 79.9\\
CosGauss-relev-pretr & 76.1 & 75.9 & 77.7 & 76.9 & 83.2 & 78.0 & 77.1 & 79.7 & \textbf{84.7} & 77.0 & 81.3 & 80.0\\
CosGauss-relev-pretr-fine& \textbf{81.2} & 77.0 & 80.9 & 78.4 & \textbf{84.9} & \textbf{80.5} &
78.3 & 81.5 & 83.5 & 78.0 & 81.5 & 80.6\\ 
\hdashline[1pt/1pt]
CosGauss-SSL-pretr & 77.3 & 73.9 & 79.6 & 79.3 & 81.6 & 78.3 & 78.7 & 80.5 & 81.9 & 75.8 & 81.8 & 79.7\\
CosGauss-SSL-pretr-fine & 78.5 & \textbf{77.9} &  77.9  & \textbf{80.2 } & 81.3  & 79.2  & \textbf{80.7} & 81.7 & 82.7 & 76.8 & 78.3 & 80.0\\
\hline
\end{tabular}

\label{tab:results}
\end{table*}

\subsection{Relevance Weighting}
The generated spectrogram of dimension $F\times T$ is passed to a self-attention-based neural network consisting of two layers of a feed forward network. For each time-frequency bin ($t,f$), we input a vector of $102$ dimensions obtained for the $i$-th filterbank representation with $\pm 51$ frame window on either side of the current frame. The output of the relevance network is a scalar that is passed through a sigmoid function to generate a relevance weight for the ($t,f$) bin. The relevance weights generated for all the ($t,f$) bins (denoted as $\boldsymbol{M}$ in Figure~\ref{fig:pipeline}) are then element-wise multiplied with the learned representation $\boldsymbol{I}$ to generate the relevance weighted ($t,f$) representation ($\boldsymbol{J}=\boldsymbol{I} \otimes \boldsymbol{M}$). 
This final representation $\boldsymbol{J}$ is used in the subsequent model for the task for COVID-19 detection. 

\begin{figure}[t!]
    \centering
    \includegraphics[width=.45\textwidth]{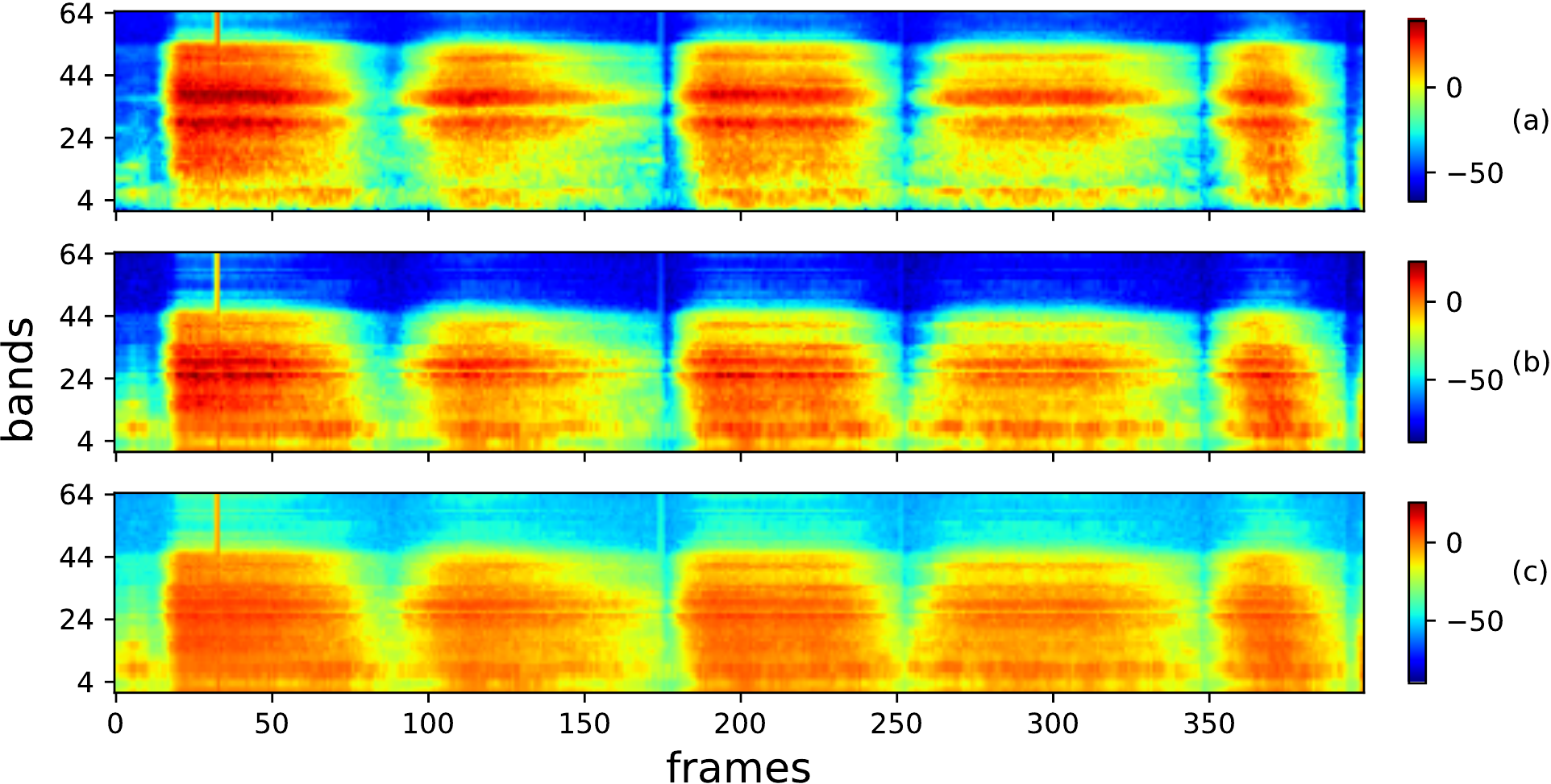}
    \caption{(a) Mel spectrogram, (b) the learned spectrogram and (c) relevance weighted spectrogram of a breathing sample.} 
    \label{fig:spectrograms}
\end{figure}
The spectrogram of a breathing signal, generated using the mel-scale filterbank and the learned filters, are shown in Figure~\ref{fig:spectrograms} (a) and (b), respectively.  
We can observe that the model distributes the filters such that low- and high-frequency bands are more stretched, compared to the mel spectrogram, to provide more detailed representation of those regions.
%The relevance mask with which the spectrogram is point-wise multiplied is shown in Figure~\ref{fig:spectrograms} (c). Unlike the relevance mask proposed in \cite{agrawal2020interpretable} which is fixed across time, the mask in the proposed model has a dynamic structure which changes across time and frequency to provide more enhanced representation. 
The weighted spectrogram is presented in Figure~\ref{fig:spectrograms} (c). Compared with (b), we  observe that the high frequency regions
% which are more relevant for the COVID-19 detection 
are more enhanced than other regions in the relevance weighted output.
% Considering both the distribution of the learned filters and the relevance weights, we can see that the model provides a more interpretable signal representation for the COVID-19 detection task.

%\input{LaTeX/figs/fig_relev_boxplot}

\subsection{Back-end Model Architecture}
The relevance weighted representations are used to compute the delta and double-delta features. 
For the supervised training, we base the rest of the architecture design on the baseline system provided as part of the Second DiCOVA challenge \cite{sharma2021second}. The model consists of $2$ layers of a bidirectional long-short term memory (BLSTM) network followed by a two class posterior output. The model is trained using the binary cross entropy (BCE) loss\footnote{The code and models are available at: \url{https://github.com/iiscleap/acoustic_repLearn_dicova2}}.

% \subsection{Modulation filter-bank}
% \begin{itemize}[noitemsep]
%     \item Rate and scale filtering. 
%     \item Modulation kernel expression.
% \end{itemize}

% \begin{equation}
%     g_i(a\,b) = \cos{2\pi (\mu_{r_i} a \pm \mu_{s_i} b)} \times \exp{\{(-a^2)+(-b^2)\}} 
% \end{equation} 
% with a sampled at $100$ Hz (corresponding to $10$ ms hop in
% the time-frequency representation), $b$ sampled at \textcolor{red}{$24$} cyc. per octave, and $i = 1,..., K$ for $K$ modulation filters. 

\section{Transfer Learning of Representations}

\subsection{Data Resources}
To develop and evaluate the proposed method, we used two data sets: The COVID-19 Sounds~\cite{xia2021covid} and the Coswara~\cite{sharma2021second} dataset. 
\subsubsection{COVID-19 Sounds Data Set}
This data set, which is crowd-sourced from $36,116$ participants from around the world, consists of $53,449$ audio samples of three different acoustic modalities, namely speech, cough, and breathing. In addition to the audio samples, demographic information, health conditions of the participants, and participants' self-reported COVID-19 testing status are also provided.
Since the testing status is self-reported, we have selected participants who tested positive within the $14$ days of the recording as the positive class, and those who never tested positive as the negative class.
Moreover, as the samples were captured from a wide range of platforms, the sampling frequency and the audio format of the samples are diverse. Therefore, we only selected the samples of $44.1$ kHz sampling frequency and those with ``wav'' audio file format.
The selected samples are then split into training and validation subsets of $1,810$ and $440$ samples, respectively for speech modality, and $1,875$ and $469$, respectively for breathing modality.

\subsubsection{Coswara Data Set}
The Coswara data set ~\cite{sharma2005coswara} is a  crowed-sourced dataset from more than $1,500$  participants across the world.
During the data collection, participants were asked to record their voice in various modalities, namely breathing, cough, sustained vowel /a/ phonation, and speech. Along with the voice samples, demographic information and self-reported COVID-19 status are also collected. 
The samples were captured through a website application and at a sampling frequency of $48$ kHz.
For this study, we used a subset of the Coswara data set of $965$ participants split into $5$ folds of 80\%--20\% training-validation, provided by the organizing committee of the Second DiCOVA challenge ~\cite{sharma2021second}.  

\subsection{Supervised Pre-training}
In supervised pre-training, we first learn the filter on the larger data set along with the target labels. In our case, we use the COVID-19 Sounds data for the pre-training. 
% For this task, we use the class label information provided with the data. 
The filters trained using this data set are then used to initialize filters for learning on the Coswara data. 
We report results in Table \ref{tab:results} for two settings, one in which pre-trained filters are not fine-tuned (CosGauss-relev-pretr) and one in which pretrained filters are fine-tuned (CosGauss-relev-pretr-fine) using the Coswara data. 
% We observe  when allow finetuning for the pre-trained filters then there is significant improvement in the final performance. This is consistent across folds for Breathing modality. 

\subsection{Self-supervised Pre-training}

\begin{figure}[t!]
    \centering
    \includegraphics[width=.33\textwidth]{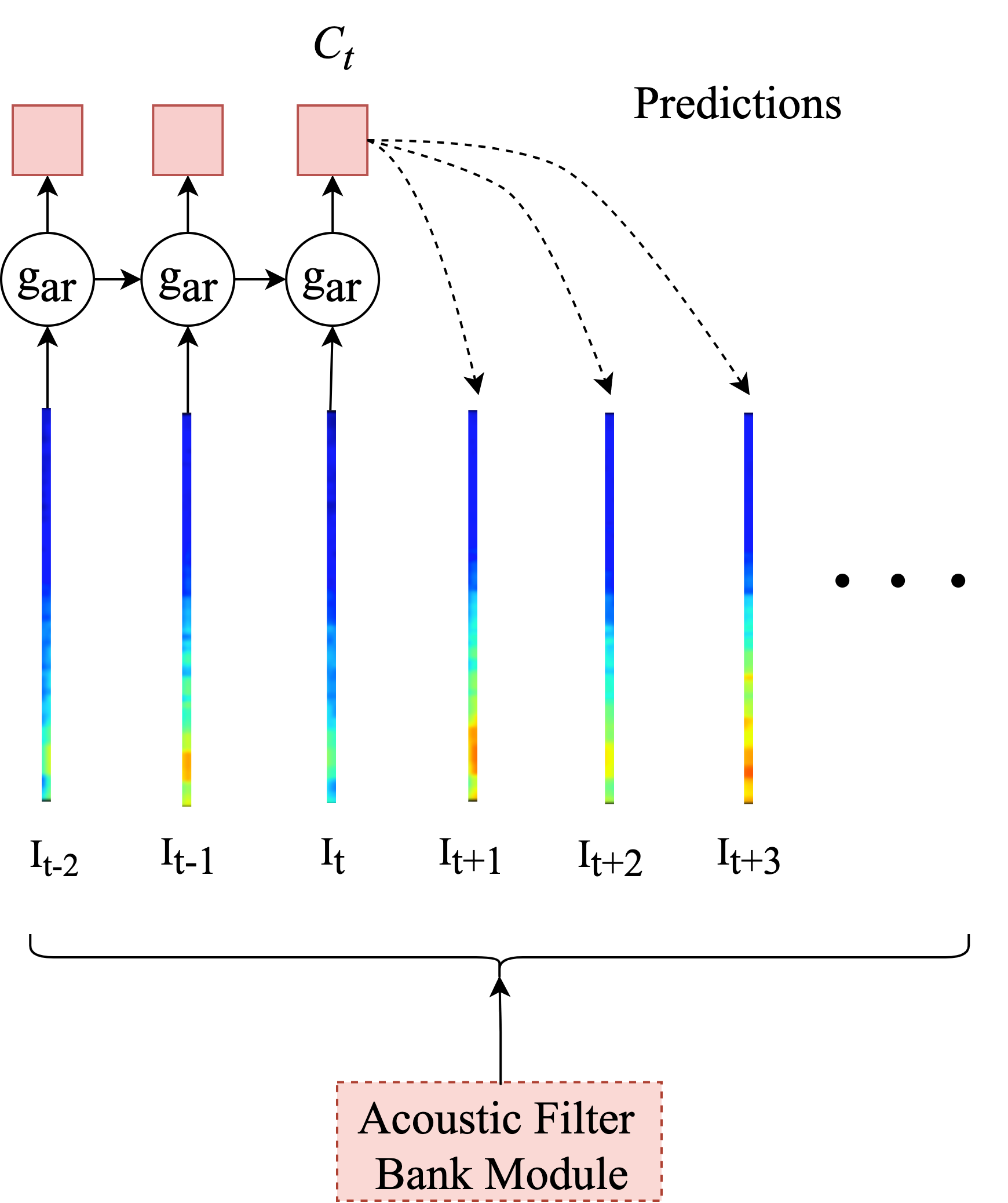}
    \caption{The proposed CPC-based architecture for self-supervised filter learning. $g_{ar}$ and $c_{t}$ represent the auto regressive module and the context vector, respectively~\cite{cpc}. The input to each $g_{ar}$ is of dimension $F\times1$. The detailed architecture of acoustic Filter Bank Module is shown in Figure \ref{fig:pipeline}.}
    \label{fig:sslcpc}
\end{figure}
%\vspace{-0.2cm}
%We also learn filters in self-supervised manner. 
For our experiments, we used the contrastive predictive coding (CPC) \cite{cpc}, which is an autoregressive method. The input audio signals are first converted into compact representations using the encoder module. Then, these representations are used to predict future representations using autoregressive models implemented as a recurrent layer.

The proposed architecture for self-supervised learning is shown in Figure \ref{fig:sslcpc}.
The acoustic filterbank module,  described in Figure \ref{fig:pipeline}, generates representations at the  frame-level of dimension $F\times1$. These are then passed through an LSTM block.
The CPC training is performed on the COVID-19 Sounds samples without using the label information. 
We hypothesize this scenario to reflect semi-supervised learning frameworks, where a large data set (in our case, the COVID-19 Sounds data set) is available without label information, while the target data set (in our case, the Coswara data set) is labelled but smaller in size. 

%Following the CPC training \cite{cpc}, the LSTM  is initialized with a $256$ dimensional hidden state. All the hyperparameters and loss functions are similar to \cite{cpc}. %

We used the COVID-19 Sounds data set to pre-train the filters in self-supervised manner. Then, the pre-trained filters are used to initialize filters in our model which are then fine-tuned on the Coswara data. 
% We report performance numbers of two experiments: when pre-trained filters are allowed to fine-tune, and when we do not fine-tune the self-supervised learning-based filters.

\begin{figure}[t!]
    \centering
    \includegraphics[width=.45\textwidth]{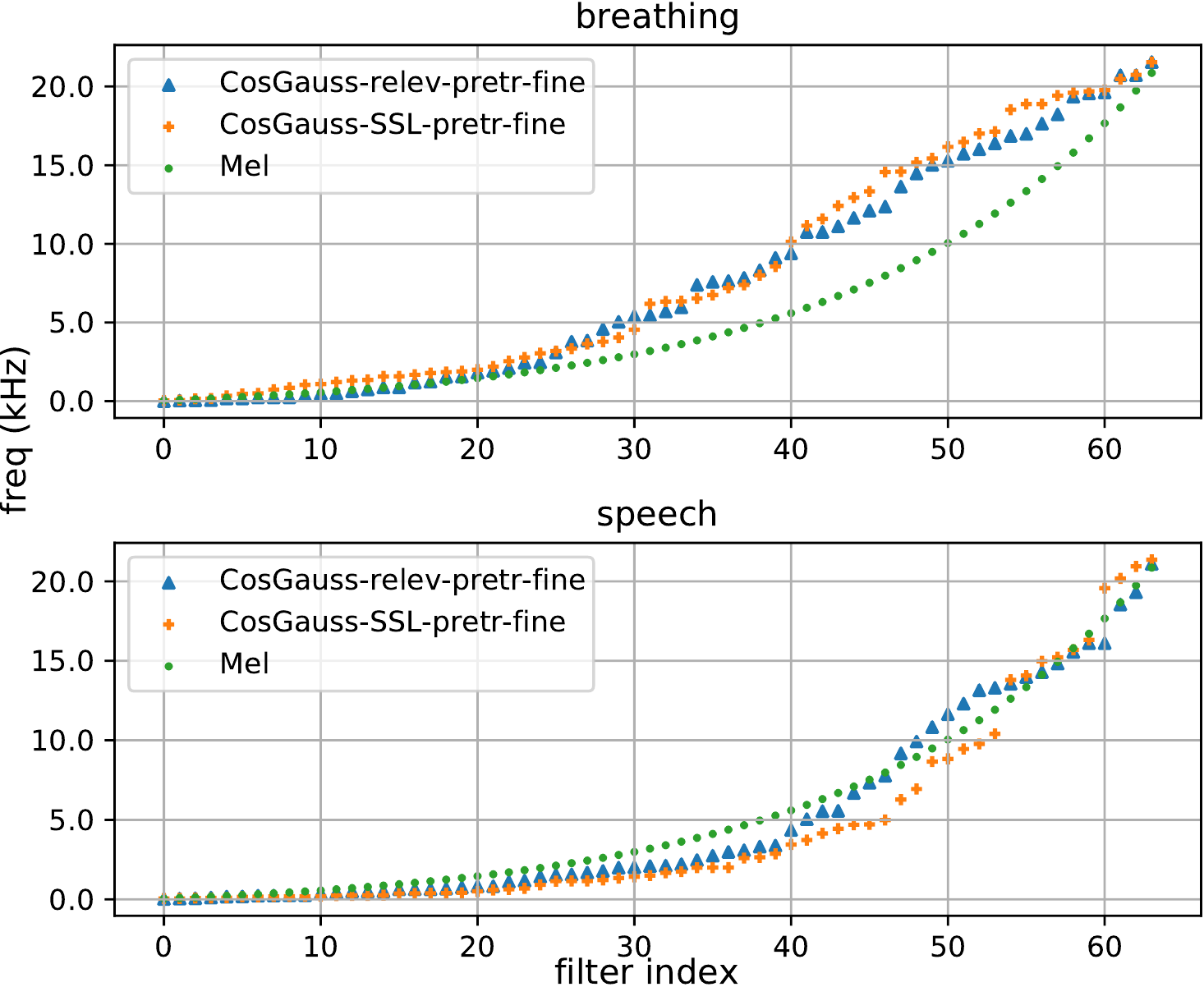}
    \vspace{-2mm}
    \caption{Distribution of centre frequencies of mel and CosGauss filters learned from Coswara data with pre-training in a supervised (CosGauss-relev-pretr-fine) and self-supervised (CosGauss-SSL-pretr-fine) fashion.} 
    \label{fig:center_freq}
    \vspace{-1mm}
\end{figure}

% \begin{itemize}[noitemsep]
%     \item How Gaussian filter front-end is incorporated in CPC encoder. 
%     \item Joint training of filters with contrastive loss.
%     \item Transfer/tune learnt filters to covid-19 prediction (dicova2) task.
% \end{itemize}

\section{Results}
Table~\ref{tab:results} compares the performance of different signal representations on breathing and speech modalities in terms of the area under the receiver operating characteristic curve (AUC). 
The representation learning methods above the dotted line are the baseline approaches compared in this paper. 
The first row shows the mel-spectrogram representation, followed by the SincNet representations \cite{ravanelli2018interpretable} in the second row.  
The third row shows the performance of LEAF representations~\cite{zeghidour2021leaf}. 

The fourth row shows the relevance weighting-based representation using cosine modulated Gaussian filters and relevance weighting.
These results improve over the baseline approaches (absolute improvements of $1$\% in AUC for breathing modality and $1.9$\% in the case of speech modality) for both the modalities of speech and breathing. Further, the improvements are consistent on four of the five folds in speech and three of the five folds in breathing.  

The fifth and sixth rows show the results for supervised pre-training without and with fine-tuning, respectively.
The results for self-supervised pre-training without and with fine-tuning are represented in the seventh and eighth rows, respectively.
These results represent the various transfer learning approaches explored in this work. The following are the major insights derived from these experiments,

\begin{itemize}[leftmargin=*]
    \item The results with pre-training on the COVID-19 Sounds data set alone in supervised (fifth row) or self-supervised fashion (seventh row) without fine-tuning degrades the performance over the direct supervised learning of the filters on the Coswara data set (fourth row) for both the speech and breathing modalities.
    \item Using the fine-tuning followed the pre-training improves the performance of the transfer learning. The supervised pre-training with fine-tuning (sixth row) offers improved performance over the self-supervised pre-training (eight row) for speech and breathing modalities.
    \item For the setting of supervised pre-training with fine-tuning, the transfer learning improves significantly (absolute improvements of $2.1$\%) in the case of breathing modality while the transfer learning approaches are inferior to the one trained directly on the Coswara data set for the speech modality. We hypothesize that speech signals are more cultural and linguistic in nature, where the speech modality of COVID-19 Sounds data set, collected primarily in UK, differs drastically from the Coswara data set collected in the Indian sub-continent. On the other hand, the breathing signals are more universal and less linguistically and geographically influenced. Hence, the representations learned on the breathing signals in the COVID-19 Sounds data set transfer well to the Coswara data set.    
\end{itemize}

\section{Discussion}
Figure~\ref{fig:center_freq} gives a better insight on how the center frequency of the filters are distributed for the different learning approaches explored in this work. 
We can observe that, for breathing, the distribution of learned filters in the low frequency regions (up to $\sim$2 kHz) is almost similar to that of the mel-scale filterbank. However, they take on a different distribution in the mid-range and high frequency range. Compared to the mel filterbank, the model puts fewer filters in the frequency regions that are not very informative for classifier and puts more filters in the regions that require more frequency resolutions. %We can see more gaps between the clusters of filters in the fine-tuned model compared to the pre-trained one.
The filter profile of the self-supervised model is similar to the supervised models in the low frequency regions. However, unlike supervised models, it places more filters in the mid-frequency range for speech data. It is interesting that the filters of the self-supervised model that are fine-tuned on the target data, take on an almost similar distribution to that of the supervised model.
For the speech, the model puts more filters than the mel filterbank in the low frequency region (0-4 kHz). Then, it becomes more sparse in the higher frequency regions.
% Figure~\ref{fig:relev_box} illustrates the relevance weights across different frequency bands. It indicates which time-frequency bins should be more emphasized. We can observe that, on average, the model put more weights on the higher bands. 
% This is in consistent with the filter profiles shown in Figure~\ref{fig:center_freq} in which the model puts more filters in the lower frequency regions and fewer filters in the mid-range and higher frequency bands. 
% This implies that this region provides more information for the classifier. Another interesting observation is that the positive and negative classes have different weights in frequency bands from 13 up to 42, which can imply that most of the voice symptoms caused by the COVID-19 are expected to manifest in those frequency bands.

\section{Conclusions}
In this paper, we have presented an approach for representation learning based on a parameterized convolutional neural network layer with a relevance weighting. The approach uses cosine modulated Gaussian kernels to learn the sub-band decomposition of the audio signal. Further, we have explored the transfer learning capabilities of the front-end using a supervised and self-supervised modeling framework. 
In our experiments on COVID-19 detection task from raw audio, we show that the proposed approach improves over the baseline mel spectrogram representations as well as the other approaches for representation learning. The improvements are also observed for both the modalities of speech and breathing, indicating the usefulness of the modeling paradigm for a broad range of acoustic signals.

%\section{Acknowledgements}

\bibliographystyle{IEEEtran}

\bibliography{mybib}

\begin{thebibliography}{10}
\providecommand{\url}[1]{#1}
\csname url@samestyle\endcsname
\providecommand{\newblock}{\relax}
\providecommand{\bibinfo}[2]{#2}
\providecommand{\BIBentrySTDinterwordspacing}{\spaceskip=0pt\relax}
\providecommand{\BIBentryALTinterwordstretchfactor}{4}
\providecommand{\BIBentryALTinterwordspacing}{\spaceskip=\fontdimen2\font plus
\BIBentryALTinterwordstretchfactor\fontdimen3\font minus
  \fontdimen4\font\relax}
\providecommand{\BIBforeignlanguage}[2]{{%
\expandafter\ifx\csname l@#1\endcsname\relax
\typeout{** WARNING: IEEEtran.bst: No hyphenation pattern has been}%
\typeout{** loaded for the language `#1'. Using the pattern for}%
\typeout{** the default language instead.}%
\else
\language=\csname l@#1\endcsname
\fi
#2}}
\providecommand{\BIBdecl}{\relax}
\BIBdecl

\bibitem{fagherazzi2021voice}
G.~Fagherazzi, A.~Fischer, M.~Ismael, and V.~Despotovic, ``Voice for health:
  the use of vocal biomarkers from research to clinical practice,''
  \emph{Digital biomarkers}, vol.~5, no.~1, pp. 78--88, 2021.

\bibitem{avila2021investigating}
F.~Avila, A.~H. Poorjam, D.~Mittal, C.~Dognin, A.~Muguli, R.~Kumar, S.~R.
  Chetupalli, S.~Ganapathy, and M.~Singh, ``Investigating feature selection and
  explainability for {COVID}-19 diagnostics from cough sounds,'' in
  \emph{INTERSPEECH}, vol.~6, 2021, pp. 4246--4250.

\bibitem{holzinger2018machine}
A.~Holzinger, ``From machine learning to explainable {AI},'' in \emph{2018
  world symposium on digital intelligence for systems and machines
  (DISA)}.\hskip 1em plus 0.5em minus 0.4em\relax IEEE, 2018, pp. 55--66.

\bibitem{low2020uncovering}
D.~M. Low, G.~Randolph, V.~Rao, S.~S. Ghosh, and P.~C. Song, ``Uncovering the
  important acoustic features for detecting vocal fold paralysis with
  explainable machine learning,'' \emph{medRxiv}, 2020.

\bibitem{gupta2016pathological}
R.~Gupta, T.~Chaspari, J.~Kim, N.~Kumar, D.~Bone, and S.~Narayanan,
  ``Pathological speech processing: State-of-the-art, current challenges, and
  future directions,'' in \emph{IEEE International Conference on Acoustics,
  Speech and Signal Processing (ICASSP)}, 2016, pp. 6470--6474.

\bibitem{ozdas2004investigation}
A.~Ozdas, R.~G. Shiavi, S.~E. Silverman, M.~K. Silverman, and D.~M. Wilkes,
  ``Investigation of vocal jitter and glottal flow spectrum as possible cues
  for depression and near-term suicidal risk,'' \emph{IEEE transactions on
  Biomedical engineering}, vol.~51, no.~9, pp. 1530--1540, 2004.

\bibitem{alghowinem2013detecting}
S.~Alghowinem, R.~Goecke, M.~Wagner, J.~Epps, M.~Breakspear, and G.~Parker,
  ``Detecting depression: a comparison between spontaneous and read speech,''
  in \emph{2013 IEEE International Conference on Acoustics, Speech and Signal
  Processing}.\hskip 1em plus 0.5em minus 0.4em\relax IEEE, 2013, pp.
  7547--7551.

\bibitem{shama2006study}
K.~Shama, A.~Krishna, and N.~U. Cholayya, ``Study of harmonics-to-noise ratio
  and critical-band energy spectrum of speech as acoustic indicators of
  laryngeal and voice pathology,'' \emph{EURASIP Journal on Advances in Signal
  Processing}, vol. 2007, pp. 1--9, 2006.

\bibitem{maslan2011maximum}
J.~Maslan, X.~Leng, C.~Rees, D.~Blalock, and S.~G. Butler, ``Maximum phonation
  time in healthy older adults,'' \emph{Journal of Voice}, vol.~25, no.~6, pp.
  709--713, 2011.

\bibitem{bengio2013representation}
Y.~Bengio, A.~Courville, and P.~Vincent, ``Representation learning: A review
  and new perspectives,'' \emph{IEEE Transactions on pattern analysis and
  machine intelligence}, vol.~35, no.~8, pp. 1798--1828, 2013.

\bibitem{radford2015unsupervised}
A.~Radford, L.~Metz, and S.~Chintala, ``Unsupervised representation learning
  with deep convolutional generative adversarial networks,'' \emph{arXiv
  preprint arXiv:1511.06434}, 2015.

\bibitem{mikolov2013efficient}
T.~Mikolov, K.~Chen, G.~Corrado, and J.~Dean, ``Efficient estimation of word
  representations in vector space,'' \emph{arXiv preprint arXiv:1301.3781},
  2013.

\bibitem{turian2022hear}
J.~Turian, J.~Shier, H.~R. Khan, B.~Raj, B.~W. Schuller, C.~J. Steinmetz,
  C.~Malloy, G.~Tzanetakis, G.~Velarde, K.~McNally \emph{et~al.}, ``Hear 2021:
  Holistic evaluation of audio representations,'' \emph{arXiv preprint
  arXiv:2203.03022}, 2022.

\bibitem{baevski2020wav2vec}
A.~Baevski, Y.~Zhou, A.~Mohamed, and M.~Auli, ``wav2vec 2.0: A framework for
  self-supervised learning of speech representations,'' \emph{Advances in
  Neural Information Processing Systems}, vol.~33, pp. 12\,449--12\,460, 2020.

\bibitem{freitag2017audeep}
M.~Freitag, S.~Amiriparian, S.~Pugachevskiy, N.~Cummins, and B.~Schuller,
  ``audeep: Unsupervised learning of representations from audio with deep
  recurrent neural networks,'' \emph{The Journal of Machine Learning Research},
  vol.~18, no.~1, pp. 6340--6344, 2017.

\bibitem{sainath2013learning}
T.~N. Sainath, B.~Kingsbury, A.-r. Mohamed, and B.~Ramabhadran, ``Learning
  filter banks within a deep neural network framework,'' in \emph{IEEE workshop
  on automatic speech recognition and understanding}, 2013, pp. 297--302.

\bibitem{hoshen2015speech}
Y.~Hoshen, R.~J. Weiss, and K.~W. Wilson, ``Speech acoustic modeling from raw
  multichannel waveforms,'' in \emph{IEEE international conference on
  acoustics, speech and signal processing (ICASSP)}, 2015, pp. 4624--4628.

\bibitem{zeghidour2021leaf}
N.~Zeghidour, O.~Teboul, F.~d.~C. Quitry, and M.~Tagliasacchi, ``Leaf: A
  learnable frontend for audio classification,'' \emph{arXiv preprint
  arXiv:2101.08596}, 2021.

\bibitem{anden2014deep}
J.~And{\'e}n and S.~Mallat, ``Deep scattering spectrum,'' \emph{IEEE
  Transactions on Signal Processing}, vol.~62, no.~16, pp. 4114--4128, 2014.

\bibitem{ravanelli2018interpretable}
M.~Ravanelli and Y.~Bengio, ``Interpretable convolutional filters with
  sincnet,'' \emph{arXiv preprint arXiv:1811.09725}, 2018.

\bibitem{agrawal2017unsupervised}
P.~Agrawal and S.~Ganapathy, ``Unsupervised modulation filter learning for
  noise-robust speech recognition,'' \emph{The Journal of the Acoustical
  Society of America}, vol. 142, no.~3, pp. 1686--1692, 2017.

\bibitem{sailor2016filterbank}
H.~B. Sailor and H.~A. Patil, ``Filterbank learning using convolutional
  restricted boltzmann machine for speech recognition,'' in \emph{IEEE
  international conference on acoustics, speech and signal processing
  (ICASSP)}, 2016, pp. 5895--5899.

\bibitem{agrawal2019unsupervised}
P.~Agrawal and S.~Ganapathy, ``Unsupervised raw waveform representation
  learning for {ASR}.'' in \emph{INTERSPEECH}, 2019, pp. 3451--3455.

\bibitem{schneider2019wav2vec}
S.~Schneider, A.~Baevski, R.~Collobert, and M.~Auli, ``wav2vec: Unsupervised
  pre-training for speech recognition,'' \emph{arXiv preprint
  arXiv:1904.05862}, 2019.

\bibitem{agrawal2018comparison}
P.~Agrawal and S.~Ganapathy, ``Comparison of unsupervised modulation filter
  learning methods for asr.'' in \emph{INTERSPEECH}, 2018, pp. 2908--2912.

\bibitem{pascual2019learning}
S.~Pascual, M.~Ravanelli, J.~Serra, A.~Bonafonte, and Y.~Bengio, ``Learning
  problem-agnostic speech representations from multiple self-supervised
  tasks,'' \emph{arXiv preprint arXiv:1904.03416}, 2019.

\bibitem{agrawal2020interpretable}
P.~Agrawal and S.~Ganapathy, ``Interpretable representation learning for speech
  and audio signals based on relevance weighting,'' \emph{IEEE/ACM Transactions
  on Audio, Speech, and Language Processing}, vol.~28, pp. 2823--2836, 2020.

\bibitem{sharma2021second}
N.~K. Sharma, S.~R. Chetupalli, D.~Bhattacharya, D.~Dutta, P.~Mote, and
  S.~Ganapathy, ``The second dicova challenge: Dataset and performance analysis
  for covid-19 diagnosis using acoustics,'' \emph{arXiv preprint
  arXiv:2110.01177}, 2021.

\bibitem{xia2021covid}
T.~Xia, D.~Spathis, J.~Ch, A.~Grammenos, J.~Han, A.~Hasthanasombat,
  E.~Bondareva, T.~Dang, A.~Floto, P.~Cicuta \emph{et~al.}, ``Covid-19 sounds:
  A large-scale audio dataset for digital respiratory screening,'' in
  \emph{Thirty-fifth Conference on Neural Information Processing Systems
  Datasets and Benchmarks Track (Round 2)}, 2021.

\bibitem{sharma2005coswara}
N.~Sharma, P.~Krishnan, R.~Kumar, S.~Ramoji, S.~Chetupalli, P.~Ghosh, and
  S.~Ganapathy, ``Coswara—a database of breathing, cough, and voice sounds
  for covid-19 diagnosis. arxiv 2020,'' \emph{arXiv preprint arXiv:2005.10548}.

\bibitem{cpc}
A.~v.~d. Oord, Y.~Li, and O.~Vinyals, ``Representation learning with
  contrastive predictive coding,'' \emph{arXiv preprint arXiv:1807.03748},
  2018.

\end{thebibliography}

% \begin{thebibliography}{9}
% \bibitem[1]{Davis80-COP}
%   S.\ B.\ Davis and P.\ Mermelstein,
%   ``Comparison of parametric representation for monosyllabic word recognition in continuously spoken sentences,''
%   \textit{IEEE Transactions on Acoustics, Speech and Signal Processing}, vol.~28, no.~4, pp.~357--366, 1980.
% \bibitem[2]{Rabiner89-ATO}
%   L.\ R.\ Rabiner,
%   ``A tutorial on hidden Markov models and selected applications in speech recognition,''
%   \textit{Proceedings of the IEEE}, vol.~77, no.~2, pp.~257-286, 1989.
% \bibitem[3]{Hastie09-TEO}
%   T.\ Hastie, R.\ Tibshirani, and J.\ Friedman,
%   \textit{The Elements of Statistical Learning -- Data Mining, Inference, and Prediction}.
%   New York: Springer, 2009.
% \bibitem[4]{YourName17-XXX}
%   F.\ Lastname1, F.\ Lastname2, and F.\ Lastname3,
%   ``Title of your INTERSPEECH 2022 publication,''
%   in \textit{Interspeech 2022 -- 23\textsuperscript{rd} Annual Conference of the International Speech Communication Association, September 18-22, Incheon, Korea, Proceedings, Proceedings}, 2022, pp.~100--104.
% \end{thebibliography}

\end{document}